\documentclass[preprint,11pt]{elsarticle}

\makeatletter
\def\ps@pprintTitle{%
  \let\@oddhead\@empty
  \let\@evenhead\@empty
  \def\@oddfoot{\reset@font\hfil\thepage\hfil}
  \let\@evenfoot\@oddfoot
}
\makeatother

\usepackage{amsmath,amsthm,amssymb,color,graphicx}
\usepackage{amscd,verbatim}
\usepackage[all]{xy}
\usepackage[breaklinks=true,colorlinks=true,linkcolor=blue,urlcolor=blue,citecolor=blue]{hyperref}
\usepackage[top=1.4in, bottom=1.4in, left=1.5in, right=1.5in]{geometry}
\newcommand{\im}{{\rm i}}

\journal{}

\begin{document}

\begin{frontmatter}



\title{T-Duality of Topological Insulators}



\author{Varghese Mathai$^{1}$  and Guo Chuan Thiang$^{1}$}

\address{
$^1$School of Mathematical Sciences, University of Adelaide,
Adelaide, SA 5005, Australia}

\begin{abstract}Topological insulators and D-brane charges in string theory can both be classified by the same family of groups. In this paper, we extend this connection via a geometric transform, giving a novel duality of topological insulators which can be viewed as a condensed matter analog of T-duality in string theory. For 2D Chern insulators, this duality exchanges the rank and Chern number of the valence bands.
\end{abstract}

\begin{keyword}
\PACS{11.25.Tq, 02.40.Re, 03.65.Vf, 73.20.At}
\end{keyword}

\end{frontmatter}

Topological insulators are experimentally observed materials that have the unusual property that
they behave like insulators in the bulk but have topologically protected conducting edge states on the boundary \cite{HK,FC}.
The topological
protection partly arises from generic symmetries of the underlying one-particle Hamiltonians that include time-reversal $T$ and charge-conjugation or particle-hole symmetry $C$
\cite{AK,RSFL,FM,T}.
Interestingly, it has been observed for some time that D-branes (charges) in orientifolds,
as well as Hamiltonians arising from topological insulators, are classified by the same 
family of groups known as K-theory, cf. \cite{RT} and Table \ref{table:0dgappedphasestable}. Although this remains just an observation, in this paper we extend it as follows. A geometric analog of the Fourier transform,
known as the Fourier--Mukai (FM) transform was used in \cite{H} to establish T-duality in 
string theory. For the condensed matter analog of S-duality in string theory, see \cite{Karch}. We utilise the FM-transform here to establish a novel duality for topological insulators. This has the interesting effect of interchanging the rank and Chern number for Chern insulators in 2D --- in particular, these two topological invariants may be considered to have equal status in our approach.

The electronic bands of insulators in $d$ spatial dimensions are usually modelled as vector bundles (or \emph{Bloch bundles}) $\mathcal{E}$ over the Brillouin torus $\mathbb{T}^d$. Such bundles arise, e.g., in the Bloch--Floquet decomposition of $\mathbb{Z}^d$ translation-invariant Hamiltonians $H$ over the unitary characters $\mathbb{T}^d$ of $\mathbb{Z}^d$ \cite{RS}. In an insulator, the Fermi level lies in a gap of the spectrum of the Hamiltonian, distinguishing as sub-bundles the conduction bands from the valence bands. Each of these sub-bundles may be robustly characterised by various topological indices.

In \cite{Hal}, Haldane introduced an example of what is now called a \emph{Chern insulator}. There, the first Chern number of the filled band in a two-band model provides a topological invariant. Like the Integer Quantum Hall Effect, it is crucial that time-reversal symmetry is absent. Topological invariants of a dramatically different nature occur in time-reversal invariant systems. This was first noticed in the seminal papers of \cite{KM,KM2}, where the authors showed that the constraint of time-reversal symmetry in their two-dimensional models leads to a $\mathbb{Z}_2$-valued index distinguishing two topologically distinct phases. Generalising to three dimensions, it was found that four $\mathbb{Z}_2$ indices could be defined for time-reversal invariant band structures \cite{FKM,MB}, three of which refer to ``weak'' topological insulators. A tantalising pattern began to emerge \cite{RSFL} and it was Kitaev who first noticed that $K$-theory and Clifford algebras provide an unifying framework for the systematic study of topological insulators \cite{AK}. This idea was greatly expanded upon in \cite{FM,T}, where the fundamental role of the underlying dynamical symmetries was emphasised. Meanwhile, it became clear that additional point symmetries result in different topological indices \cite{Fu,FK}, and that the weak insulators are actually robust and interesting in their own right \cite{RKS}. Significantly, the $K$-theoretic approach not only encompasses all the above examples, but also explains the phenomenon of dimensional shifts \cite{T} as made apparent in the Periodic Table of \cite{AK,RSFL}, where some of the classes can be realised by topological superconductors. Furthermore, it is the natural language to phrase our proposed duality in.

Such theoretical studies are particularly compelling in that they preceded several recent experiments verifying some of the predicted topological phases: the time-reversal invariant quantum spin Hall phase was realised in \cite{KW}. Then the Chern insulators were experimentally found in \cite{CZ} and \cite{JM}, the latter realising explicitly the Haldane model.

The T-duality isomorphisms which we introduce in this paper, as well as their applications, are explored in greater detail in our subsequent papers \cite{MT, MT2}. In particular, we have found that bulk-boundary maps simplify under a T-duality transformation. These examples demonstrate the conceptual and computational utility of T-duality in the analysis of topological insulator invariants.

\section{Duality for Chern insulators in 2D}\label{section:dualitychern}
To illustrate our duality concretely in a simple case without recourse to $K$-theory, we revisit the model Hamiltonians which give rise to Chern insulators. The basic example is a two-band Hamiltonian in 2D, which up to a shift in the energy bands has the generic form \cite{SPF}
\begin{equation*}
    H(k)= \mathbf{h}(k)\cdot\vec{\sigma}\equiv \sum_{i=1}^3 h_i(k)\sigma_i, \;\;k\in\mathbb{T}^2,
\end{equation*}
where $\sigma_i$ are the Pauli matrices and $\mathbf{h}=(h_1,h_2,h_3)$ is a smooth map from the Brillouin torus $\mathbb{T}^2$ to $\mathbb{R}^3$. Since $H(k)^2=|\mathbf{h}(k)|^2$, the gapped condition is ensured if $\mathbf{h}$ is nowhere zero. Assuming this, we can form the ``spectrally-flattened'' Hamiltonian $\widetilde{H}(k)=\widehat{\mathbf{h}}(k)\cdot\vec{\sigma}$ where $\widehat{\mathbf{h}}=\mathbf{h}/|\mathbf{h}|$, and the valence band is given by the projection onto its negative eigenbundle $P=\frac{1}{2}(1-\widetilde{H}(k))$. In particular, the valence band is a line bundle $\mathcal{L}$ and has the Chern number $c(\mathcal{L})$ as a topological invariant:
\begin{equation}
c(\mathcal{L})=\frac{1}{4\pi}\int_{\mathbb{T}^2}\mathrm{d}k\,\widehat{\mathbf{h}}(k)\cdot\frac{\partial\widehat{\mathbf{h}}(k)}{\partial{k_1}}\wedge\frac{\partial\widehat{\mathbf{h}}(k)}{\partial{k_2}}.\label{chernformulalinebundle}
\end{equation}

While early two-band models \cite{Hal,BHZ,QHZ} provided examples of valence bands with Chern numbers $0$ or $\pm 1$, one can in principle write down models whose valence bands have arbitrary Chern numbers by a judicious choice of the map $\mathbf{h}$ \cite{SPF}. Furthermore, there is no particular reason to restrict to two-band models. The Fermi projection $P$ onto the occupied states of a gapped Hamiltonian determines, more generally, a vector bundle $\mathcal{E}_P$ comprising the valence bands, whose Chern number is $c(\mathcal{E}_P)=\frac{i}{2\pi}\int_{\mathbb{T}^2}\mathrm{Tr}(P\mathrm{d}P\wedge \mathrm{d}P)$,
generalising \eqref{chernformulalinebundle}. Thus, two physically inequivalent situations --- a single valence band versus multiple bands --- can give rise to the same topological invariant $c$. They are of course distinguished by the rank, which is the number of filled valence bands. Although the rank $r(\mathcal{E})$ is also an honest topological invariant of a bundle $\mathcal{E}$, it is sometimes considered uninteresting and thus neglected\footnote{Note that the $\mathbb{Z}$-invariant for gapped Hamiltonians in 0-dimensions, which appears in many versions of the ``Periodic Table'' in the literature, is nothing but the Hilbert space rank.}. We aim to show, on the contrary, that the rank is dual in a precise sense to the Chern number, so both invariants should be accorded similar status. Keeping track of both $r(\mathcal{E})$ and $c(\mathcal{E})$ allows the notion of a dual topological insulator to become apparent.

Recall that the (first) Chern number $c(\mathcal{E})$ can also be understood as the integral over the Brillouin torus of the first Chern class $c_1(\mathcal{E})$; the latter can thus be represented (after normalising the total volume to $1$) by the differential 2-form $c(\mathcal{E})\mathrm{d}k_1\wedge\mathrm{d}k_2$. In general, the Chern character $\mathrm{ch}(\mathcal{E})$ of a vector bundle $\mathcal{E}$ over a base manifold $X$ is an element of the even cohomology ring of $X$. It can be expressed as
\begin{equation}
   \mathrm{ch}(\mathcal{E})=r(\mathcal{E})+c_1(\mathcal{E})+\frac{1}{2}(c_1(\mathcal{E})^2-2c_2(\mathcal{E}))+\ldots,\nonumber
\end{equation}
where each of the degree-$2i$ classes $c_i(\mathcal{E})$ can be represented by degree-$2i$ differential forms. For a valence vector bundle $\mathcal{E}$ over $\mathbb{T}^2$ with rank $r(\mathcal{E})$ and first Chern class $c_1(\mathcal{E})$, the Chern character simplifies to $\mathrm{ch}(\mathcal{E})=r(\mathcal{E})+c_1(\mathcal{E})=r(\mathcal{E})+c(\mathcal{E})\mathrm{d}k_1\wedge\mathrm{d}k_2$ and incorporates both bundle invariants $r$ and $c$.

Let $\widehat{\mathbb{T}^2}$ be another (``dual'') torus with coordinates $k_1',k_2'$ (again normalised to have volume $1$). There is a so-called \emph{Poincar\'{e} line bundle} $\mathcal{P}$ over $\mathbb{T}^2\times\widehat{\mathbb{T}^2}$, whose first Chern class can be represented by the 2-form $c_1(\mathcal{P})=\mathrm{d}k_1\wedge\mathrm{d}k_2'+\mathrm{d}k_2\wedge\mathrm{d}k_1'$. Thus its Chern character can be written as \begin{equation}
\mathrm{ch}(\mathcal{P})=1+c_1(\mathcal{P})+\frac{1}{2}c_1(\mathcal{P})\wedge c_1(\mathcal{P})=1+\mathrm{d}k_1\wedge\mathrm{d}k_2'+\mathrm{d}k_2\wedge\mathrm{d}k_1'+\mathrm{d}k_1\wedge\mathrm{d}k_2\wedge\mathrm{d}k_1'\wedge\mathrm{d}k_2'.\nonumber
\end{equation}
The \emph{Fourier--Mukai dual} (FM-dual) to $\mathcal{E}$ is a canonically given vector bundle $\widehat{\mathcal{E}}$ over $\widehat{\mathbb{T}^2}$ whose Chern character is \cite{H}
\begin{equation}
\mathrm{ch}(\widehat{\mathcal{E}})=\int_{\mathbb{T}^2}\mathrm{ch}(\mathcal{E})\mathrm{ch}(\mathcal{P}).\label{FMtwotorus}
\end{equation}
A short calculation gives $\mathrm{ch}(\widehat{\mathcal{E}})=c(\mathcal{E})+r(\mathcal{E})\mathrm{d}k_1'\wedge\mathrm{d}k_2'$, showing that $r(\widehat{\mathcal{E}})=c(\mathcal{E})$ and $c(\widehat{\mathcal{E}})=r(\mathcal{E})$, i.e., 
the transform
\eqref{FMtwotorus} {\em interchanges the rank and the Chern number}. We may interpret the dual bundle $\widehat{\mathcal{E}}$ as the valence bands of a \emph{dual topological insulator} (with Hamiltonian $\widehat{H}$) to that corresponding to $H$. Clearly, taking the dual twice brings us back to the original topological insulator.

As a non-trivial example, consider the two-band model Hamiltonians of \cite{SPF} which realise valence bands with Chern numbers $0,\pm 1$ or $\pm 2$ (alternatively, we can use the extended Haldane model described in \cite{FC}). The vector $\mathbf{h}(k)$ defining their model Hamiltonian is
\begin{equation}
\mathbf{h}(k)=(\cos k_1, \cos k_2, m+a\cos (k_1+k_2)+b(\sin k_1+\
\sin k_2)),\label{chernhamiltonian}
\end{equation}
where $m,a,b$ are real parameters. The Chern number of the valence band $\mathcal{L}$ was computed to be
\begin{equation*}
    c(\mathcal{L})=\mathrm{sgn}(-m-a)+\frac{1}{2}[\mathrm{sgn}(m-a+2b)+\mathrm{sgn}(m-a-2b)].
\end{equation*}
Taking $(m,a,b)=(1,-2,1)$ in \eqref{chernhamiltonian}, we obtain a Chern insulator whose valence band  $\mathcal{E}=\mathcal{L}$ has $(r,c)(\mathcal{E})=(1,2)$. Its FM-dual $\widehat{\mathcal{E}}$ has $(r,c)(\widehat{\mathcal{E}})=(2,1)$. We can realise $\widehat{\mathcal{E}}$ as the valence bundle of a \emph{three}-band dual Hamiltonian $\widehat{H}$, by starting with a two-band Hamiltonian $\widehat{H}_0$ defined by \eqref{chernhamiltonian} with $(m,a,b)=(1,-2,2)$, and augmenting it with a trivial valence band, i.e., $\widehat{H}=\widehat{H}_0\oplus -1$. Then we see that $\widehat{H}$ has valence bundle with rank 2 and Chern number 1, and is the dual topological insulator to $H$.

We also recall that the rank plays an important role in the Integer Quantum Hall Effect, through the enumeration of possible band gaps. Under the simplifying assumption of a rational magnetic flux per unit cell, the Landau Hamiltonian can be diagonalised in quasi-momentum space. Provided the Fermi level lies in an energy gap, the filling factor, or number of filled Landau bands, is related to the Hall conductance. Each occupied band contributes an integer (a Chern number) to the conductance according to the TKNN formula \cite{TKNN,Bel}. In this case, the rank may be identified with the integrated density of states, which is used to detect spectral gaps in the Hamiltonian \cite{Bel}.

As we will explain in Section \ref{section:dualitygeneral}, the exchange of topological invariants under duality is a generic feature of the more general Fourier--Mukai (FM) transform between bundles over Brillouin tori of arbitrary dimensions. We emphasise that the FM-transform is geometrically and canonically defined. In dimensions $d\neq 2$ and in the presence of further symmetry constraints, the duality is more complex and exposes a wealth of previously unseen structure \cite{MT, MT2}.

\section{The role of symmetry in topological insulators}
Topological $K$-theory associates to each topological space $X$ a sequence of abelian groups $K^{-n}(X), n\in\mathbb{Z}$. In the complex case, the sequence is 2-periodic. The real version comes in inter-related varieties, such as $KO$ (real), $KR$ (Real), $KSp$ (quaternionic) and $KQ$ (Quaterionic), each of which is 8-periodic. In the absence of any additional symmetry constraints (Class $A$), the complex $K^0$ group of $\mathbb{T}^d$ can be interpreted as virtual classes (i.e. formal differences) of valence bands, while the reduced group $\widetilde{K}^0(\mathbb{T}^d)$ forgets the rank and comprises their stable equivalence classes. In general, we may want to keep track of the conduction bands as well \cite{FM}. Then $K^0(\mathbb{T}^d)$ can be understood as a group classifying the obstructions in deforming one insulator into another \cite{T}. Higher-index $K$-theory groups enter whenever certain additional symmetries are also present (see below).

Symmetries of a Hamiltonian may be implemented unitarily or antiunitarily. Assuming that $0$ lies in a gap of the spectrum of $H$, we can define the grading operator $\Gamma=\textrm{sgn}(H)$. The special involutory symmetries of time-reversal $T$ and charge-conjugation $C$ may be present. They are respectively realised as antiunitary operators $\mathsf{T}$ and $\mathsf{C}$, each squaring to $\pm 1$ (since the operators only need to be involutory in a projective sense). Also, $\mathsf{T}$ is even with respect to $\Gamma$, whereas $\mathsf{C}$ is odd. It is also possible that $T$ and $C$ are not separately symmetries, but $S\equiv CT$ is. Then its representative $\mathsf{S}$ is odd, unitary, and squares to +1. To summarise, there is a $CT$-group $\{1,T\}\times\{1,C\}\cong(\mathbb{Z}_2)^2$, and each choice of $CT$-subgroup $A$ along with the squares $\mathsf{C}^2=\pm 1$ and $\mathsf{T}^2=\pm 1$ determines one of ten symmetry classes (see Table \ref{table:0dgappedphasestable} and \cite{RSFL,FM}).

In $d>0$, compatibility of $\mathsf{T},\mathsf{C}$ with $H$,
\begin{equation}
\mathsf{T}H(k)\mathsf{T}^{-1}=H(-k),\;\;\mathsf{C}H(k)\mathsf{C}^{-1}=-H(-k),\label{CTcommutation}
\end{equation}
imposes some extra structure on the admissible Bloch bundles $\mathcal{E}\rightarrow\mathbb{T}^d$. For example, if $\mathsf{T}^2=+1$, then $\mathsf{T}$ is an antilinear involution on $\mathcal{E}$ which covers the inversion $\varsigma:k\mapsto -k$ on the Brillouin torus. The inversion $\varsigma$ arises from the complex-conjugation due to $\mathsf{T}$, which takes a unitary character $\mathbb{Z}^d\ni x\mapsto e^{\im k x}$ to its complex conjugate character $x\mapsto e^{-\im k x}$. Mathematically, $\mathsf{T}$ defines a \emph{Real structure} on $\mathcal{E}$, with the consequence that $KR$-theory \cite{Ati} is the correct variant of real K-theory for the classification of topological insulators with such additional symmetries. The Real vector bundles in $KR$-theory are complex vector bundles over spaces $X$ with involutions $\varsigma$, such as $(\mathbb{T}^d;\varsigma:k\mapsto -k)$ in the previous example. A Real bundle comes with an antilinear lift $\tilde{\varsigma}$ (the Real structure) of $\varsigma$ which squares to the identity. If $\varsigma$ is trivial, the $KR$-theory of $(X;\varsigma=\mathrm{id})$ reduces to the ordinary real $KO$-theory of $X$. It is sometimes convenient to use $KQ$-theory \cite{Dup}, which involves Quaternionic bundles. The latter are similar to Real bundles in that they are complex vector bundles over spaces with involution $(X;\varsigma)$, but they are equipped with a \emph{Quaternionic structure}, i.e.\ an antilinear lift $\tilde{\varsigma}$ of $\varsigma$ such that $\tilde{\varsigma}^2=-1$, rather than a Real structure. For example, the standard fermionic time-reversal $\mathsf{T}$ has $\mathsf{T}^2=-1$, so it defines a Quaternionic structure on $\mathcal{E}$. If the involution $\varsigma$ on the base space is trivial, $KQ$-theory reduces to $KSp$-theory, the $K$-theory for symplectic vector bundles.

The Clifford algebras $Cl_{r,s}$ and $\mathbb{C}l_n$\footnote{$Cl_{r,s}$ is the real Clifford algebra generated by anticommuting elements $e_1,\ldots,e_r,\epsilon_1,\ldots\epsilon_s$, with $e_i^2=-1, \epsilon_j^2=+1$; $\mathbb{C}l_n$ is the complex Clifford algebra on $n$ generators.} are a convenient device for encoding the presence/absence of and the squares of $\mathsf{T},\mathsf{C},\mathsf{S}$. They arise from an analysis of the commutation relations amongst $\{i,\mathsf{T},\mathsf{C},\Gamma\}$ (see \cite{FM,T} for details). The general idea is that the $\mathsf{T},\mathsf{C},\mathsf{S}$ operators, together with the compatible spectrally-flattened gapped Hamiltonian $\Gamma$, equivalently define a graded action of some Clifford algebra on the representation space. As explained in \cite{Kar,T}, higher-index $K$-theory groups can be defined in terms of gradings compatible with appropriate Clifford algebra actions. In $d=0$, the relevant $K$-theory groups are listed in Table \ref{table:0dgappedphasestable}, while in $d>0$, we should replace the groups $KO^{-n}(\star)$ (the $KO$-theory groups of a point) by $KR^{-n}(\mathbb{T}^d)$, and also $K^{-n}(\star)$ by $K^{-n}(\mathbb{T}^d)$.

\begin{table}
\begin{center}
\begin{tabular}{l | l | l |c c |  c  c}
	$n$	&	\parbox[c]{1cm}{Class} & \parbox[c]{1.2cm}{Symm.} & $\mathsf{C}^2$ & $\mathsf{T}^2$  & \parbox[c]{1.7cm}{$Cl_{0,n}$\,/ $\mathbb{C}l_n$}	&	\parbox[l]{2.9cm}{$KO^{-n}(\star)$\,/ $K^{-n}(\star)$}\\
	\hline
	$0$	&	$AI$	&	$T$	    &	 	& $+1$	& $Cl_{0,0}$	& $\mathbb{Z}$ \\
	$1$	&	$BDI$	&	$C,T$ 	& $+1$  & $+1$	& $Cl_{0,1}$ 	& $\mathbb{Z}_2$ \\	
	$2$	&	$D$		&	$C$		& $+1$  &		& $Cl_{0,2}$ 	& $\mathbb{Z}_2$ \\	
	$3$	&	$DIII$	&	$C,T$ 	& $+1$  & $-1$	& $Cl_{0,3}$ 	& $0$ \\	
	$4$	&	$AII$	&	$T$		&    	& $-1$	& $Cl_{0,4}$ 	& $\mathbb{Z}$ \\	
	$5$	&	$CII$	&	$C,T$ 	& $-1$ 	& $-1$	& $Cl_{0,5}$ 	& $0$ \\	
	$6$	&	$C$		&	$C$		& $-1$ 	&		& $Cl_{0,6}$ 	& $0$ \\	
	$7$	&	$CI$	&	$C,T$ 	& $-1$ 	& $+1$	& $Cl_{0,7}$	& $0$ \\
	
	\hline\hline
	$0$	&	$A$		&	N/A		& \multicolumn{2}{c|}{N/A}   	& $\mathbb{C}l_0$ & $\mathbb{Z}$ \\
	$1$	&	$AIII$	&	$S$	& \multicolumn{2}{c|}{$\mathsf{S}^2=+1$}		& $\mathbb{C}l_1$ & $0$ \\
\end{tabular}
\caption{Classification of $0$-dimensional gapped topological phases according to their $CT$-symmetries \cite{AK,RSFL}.}
\label{table:0dgappedphasestable}
\end{center}
\end{table}

For the Fourier--Mukai transform later, a geometric picture of $C$ and/or $T$-invariant topological insulators in terms of Real or Quaternionic bundles is convenient. The groups $KR^{-n}(\mathbb{T}^d;\varsigma)$ (resp.\ $KQ^{-n}(\mathbb{T}^d;\varsigma)$) can be described using graded Real (resp.\ Quaternionic) bundles $\mathcal{E}\rightarrow\mathbb{T}^d$ with commuting graded complex-linear $Cl_{0,n}$-module structure on each fibre compatible with the Real (resp. Quaternionic) structure $\tilde{\varsigma}$ on $\mathcal{E}$. Physically, a spectrally-flattened gapped Hamiltonian $\Gamma=\mathrm{sgn}(H)$ provides a grading on the Bloch bundle, and a compatible graded $Cl_{0,n}$-action means that there are $n$ mutually anticommuting (fibrewise) operators which are odd with respect to $\Gamma=\mathrm{sgn}(H)$. Note that there is a graded Morita equivalence allowing us to convert between graded $Cl_{r,s}$-actions and $Cl_{0,s-r\,\mathrm{(mod\,\,8})}$-actions; also, a graded $Cl_{0,n}$-action is equivalently an ungraded $Cl_{0,n+1}$-action \cite{Kar, LM}. 

The two complex cases are easy to analyse. When present, the operator $\mathsf{S}$ anticommutes with $\Gamma$, so the Bloch bundles host a graded $\mathbb{C}l_1$-action and are described by $K^{-1}(\mathbb{T}^d)$. Otherwise, the graded Bloch bundles are simply classified by $K^0(\mathbb{T}^d)$. For each of the remaining eight real cases, we will identify below the odd operators generating the (graded) $Cl_{q,p}$-action, as well as the associated $KR$-theory groups.

{\bf $\mathsf{T}^2=\pm 1$, no $\mathsf{C}$.}
Use $\mathsf{T}$ as the Real or Quaternionic structure $\tilde{\varsigma}$ according to $\mathsf{T}^2=\pm 1$. The corresponding $K$-theory groups are $KR^0(\mathbb{T}^d)$ and $KQ^0(\mathbb{T}^d)\cong KR^{-4}(\mathbb{T}^d)$.

{\bf $\mathsf{T}^2=\pm 1$, $\mathsf{C}^2=\pm 1$.}
As above, use $\mathsf{T}$ as the Real/Quaternionic structure. Then $\mathsf{C}\mathsf{T}$ is odd with respect to $\Gamma$, giving either a graded $Cl_{1,0}$ action or a graded $Cl_{0,1}$-action on each fibre commuting with $\tilde{\varsigma}$, depending on the value of $(\mathsf{C}\mathsf{T})^2$. Thus for $\mathsf{T}^2=+1,\mathsf{C}^2=+1$, the Real Bloch bundles compatible with the graded $Cl_{0,1}$-action are described by $KR^{-1}(\mathbb{T}^d)$, whereas for $\mathsf{T}^2=+1,\mathsf{C}^2=-1$, we get $KR^{-7}(\mathbb{T}^d)$. Similarly, for $\mathsf{T}^2=-1,\mathsf{C}^2=+1$, the graded Quaternionic Bloch bundles compatible with the graded $Cl_{1,0}$-action are described by $KQ^{-7}(\mathbb{T}^d)\cong KR^{-3}(\mathbb{T}^d)$, whereas for $\mathsf{T}^2=-1,\mathsf{C}^2=-1$, we get $KQ^{-1}(\mathbb{T}^d)\cong KR^{-5}(\mathbb{T}^d)$.

{\bf$\mathsf{C}^2=\pm 1$, no $\mathsf{T}$.}
Use $\mathsf{C}$ as the Real/Quaternionic structure. The operator $i\Gamma$ gives an \emph{ungraded} $Cl_{1,0}$-action on the fibres. Using the ungraded Morita equivalence $Cl_{1,0}\sim Cl_{0,3}$, we can construct a graded $Cl_{0,2}$-action; the relevant $K$-theory group is then $KR^{-2}(\mathbb{T}^d)$ for $\mathsf{C}^2=+1$, and $KQ^{-2}(\mathbb{T}^d)\cong KR^{-6}(\mathbb{T}^d)$ for $\mathsf{C}^2=-1$.

\begin{figure}[ht]
\begin{center}
\includegraphics{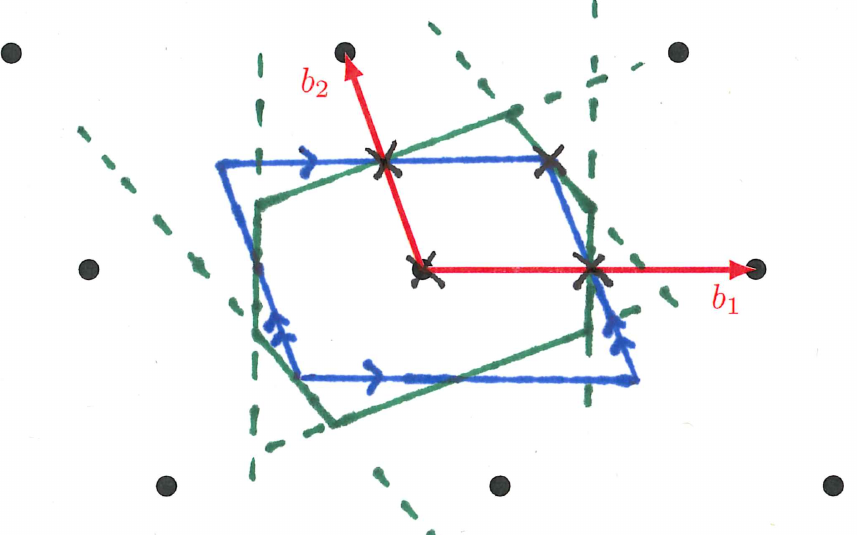}\\
\end{center}
\caption{A generic 2D reciprocal lattice. The quasi-momentum $k$ is defined up to translations generated by vectors $b_1,b_2$, and the solid green lines bound the Brillouin zone. It is topologically a torus (blue parallelogram). Each of $\mathsf{T},\mathsf{C},\mathsf{P}$ induces the same inversion $\varsigma:k\mapsto -k$, and the crosses mark the four fixed points. The fibre above such a point hosts a real/quaternionic structure if $C$ and/or $T$ is present, and a representation of $\mathbb{Z}_2=\{1,P\}$ if $P$ is present. They play an important role in the calculation of some topological invariants \cite{FM, FK2}.}
\label{fig:brillouin}
\end{figure}

\section{Discussion and $P$-symmetry}
The appearance of $KR$-theory groups appears to make the $K$-theory classification rather complicated. Some simplifications are often made in the literature, which has the effect of reducing attention to the $K$-theory groups of a point. One such approach dispenses with the Bloch-theory description altogether, and restricts attention to Dirac-type Hamiltonians. The classification problem becomes that of finding extensions of $Cl_{r,s}$-actions to $Cl_{r,s+1}$-actions. Roughly speaking, the $Cl_{r,s}$-action comes from the symmetries $\mathsf{C},\mathsf{T}$ along with the $d$ gamma matrices in the Dirac Hamiltonian. Then one looks for compatible mass terms giving an extra Clifford generator, which may be taken to be a grading operator \cite{AK}. Although there is an \emph{algebraic} periodicity in the Clifford algebras, this is weaker than the full (Bott) periodicity in $K$-theory. A second approach ``approximates'' the Brillouin torus by the sphere $S^d$. The standard argument for this is that the ``strong'' insulators are captured within such an approximation. The full torus topology is actually much richer, and in particular, its $K$-theory contains some other groups which describe the so-called ``weak'' topological insulators. In view of the renewed interest in these weak insulators \cite{RKS}, as well as the duality theory which we will develop, we will always consider the Brillouin zone to be a torus.

Other symmetries (besides the $CT$-symmetries and translational symmetries) impose further constraints on the admissible Bloch bundles. For example, non-trivial point groups lead to \emph{crystalline} topological insulators, which must satisfy equivariance conditions. Of particular interest is spatial inversion $P$, which acts on real-space wave functions by $(P\psi)(x)=\psi(-x)$. On Bloch bundles, inversion symmetry is realised by a complex-linear involution $\mathsf{P}$ which covers the inversion $k\mapsto -k$, i.e., $\mathsf{P}H(k)\mathsf{P}^{-1}=H(-k)$, c.f.\ Fig. \ref{fig:brillouin}. In view of the fundamental status of the $P,C,T$ symmetries in quantum field theory, and the well-known special roles of $C$ and $T$ in topological insulator theory, it is natural to ask whether $P$ symmetry also has some distinguished role on the topological insulator side. We argue that this is indeed the case, illustrating the claim with a duality of ``tenfold-ways'' provided by the FM-transform.

In $d=0$, an analysis of the $CT$-subgroups suffices to produce the classification of ``topological insulators'' (or \emph{phases}) as in Table \ref{table:0dgappedphasestable}. In $d>0$, we generalise this to subgroups of the $PCT$ group $\{1,P\}\times\{1,C\}\times\{1,T\}$ rather than just the $CT$-group. Note that $P,C,T$ need not \emph{separately} be symmetries. In fact, we obtain an inequivalent tenfold-way to that in Table \ref{table:0dgappedphasestable} by analysing $\{1,CP\}\times\{1,TP\}$ and its subgroups, in analogy to our earlier analysis of $\{1,C\}\times\{1,T\}$ subgroups. Instead of \eqref{CTcommutation}, we now have
\begin{equation*}
(\mathsf{T}\mathsf{P})H(k)(\mathsf{T}\mathsf{P})^{-1}=H(k),\;\;(\mathsf{C}\mathsf{P})H(k)(\mathsf{C}\mathsf{P})^{-1}=-H(k).\label{CPTPcommutation}
\end{equation*}
For example, in a $TP$-invariant topological insulator, the antilinear operator $\mathsf{T}\mathsf{P}$ acts \emph{fibrewise} on the Bloch bundle (the involutions on $\mathbb{T}^d$ due to $\mathsf{T}$ and $\mathsf{P}$ cancel out), providing an ordinary real (as opposed to Real) structure if it squares to $+1$. This means that $KO$-theory and $KSp$-theory are the relevant variants of $K$-theory to use when some subset of $\{1,CP\}\times\{1,TP\}$ symmetries is present. There are again eight real symmetry classes, each of which has at least one of the symmetries $CP$ or $TP$.

\subsection{$PCT$-invariant topological insulators}\label{subsection:PCTinsulators}
{\bf $(\mathsf{T}\mathsf{P})^2=\pm 1$, no $\mathsf{C}\mathsf{P}$.}
A $TP$-invariant topological insulator is a graded complex vector bundle $\mathcal{E}\rightarrow\mathbb{T}^d$, equipped with an even antilinear bundle map $\mathsf{T}\mathsf{P}$ \emph{commuting} with the bundle projection. Thus $\mathsf{T}\mathsf{P}$ provides a real or quaternionic structure (as opposed to Real/Quaternionic) on $\mathcal{E}$. Such insulators can be studied using $KO^0(\mathbb{T}^d)$ or $KSp^0(\mathbb{T}^d)\cong KO^{-4}(\mathbb{T}^d)$.

{\bf $(\mathsf{T}\mathsf{P})^2=\pm 1$, $(\mathsf{C}\mathsf{P})^2=\pm 1$.}
$\mathsf{T}\mathsf{P}$ provides a real or quaternionic structure as above, while $\mathsf{C}\mathsf{P}$ gives a graded $Cl_{0,1}$ or $Cl_{1,0}$ action on each fibre commuting with $\mathsf{T}\mathsf{P}$. Such insulators can be studied using $KO^{-1}(\mathbb{T}^d), KO^{-7}(\mathbb{T}^d), KSp^{-1}(\mathbb{T}^d)\cong KO^{-5}(\mathbb{T}^d)$, or $KSp^{-7}(\mathbb{T}^d)\cong KO^{-3}(\mathbb{T}^d)$.

{\bf $(\mathsf{C}\mathsf{P})^2=\pm 1$, no $\mathsf{T}\mathsf{P}$.}
Follow the construction for the $\mathsf{C}^2=\pm 1$, no $\mathsf{T}$ case. The relevant $K$-theory groups here are $KO^{-2}(\mathbb{T}^d)$ or $KSp^{-2}(\mathbb{T}^d)\cong KO^{-6}(\mathbb{T}^d)$.

In summary, the ordinary real $KO$-theory groups of the Brillouin torus describe topological insulators with $TP$ and/or $CP$ symmetries.

\section{Duality for topological insulators}\label{section:dualitygeneral}
The geometric analog of the Fourier transform is the Fourier--Mukai transform, which may be summarised by the commutative diagram
\begin{equation}
\xymatrix{ 
& {\mathcal P} \ar[d] & \\
 &  \mathbb{T}^d\times \widehat{\mathbb{T}^d} \ar[dl]_{p} \ar[dr]^{\widehat{p}} &  \\
\mathbb{T}^d && \widehat{\mathbb{T}^d}. 
 }\label{poincarediagram}
\end{equation}
Here, $\mathcal{P}$ is a generalisation of the Poincar\'{e} line bundle introduced in Section \ref{section:dualitychern} \cite{H}. For the real version of the FM-transform, the torus $\widehat{\mathbb{T}^d}$ has the involution $\varsigma:k\mapsto -k$ and is dual to the torus $\mathbb{T}^d$ with trivial involution. $\mathcal{P}$ has the property that the involution on $\widehat{\mathbb{T}^d}$ lifts to $\mathcal P$. The real FM-transform gives rise to canonical isomorphisms \cite{MT,H}
\begin{subequations}
\begin{align}
KO^{-n+d}(\mathbb{T}^d)\cong & KR^{-n}(\widehat{\mathbb{T}^d}),\label{FM1}\\
KSp^{-n+d}(\mathbb{T}^d)\cong & KQ^{-n}(\widehat{\mathbb{T}^d}),\label{FM2} 
\end{align}
\end{subequations}
given by the formula $\widehat{p}_*(p^*(\mathcal E)\otimes \mathcal P)$ \cite{MT,H}, where $\mathcal E$ is a graded vector bundle in $KO^{-n+d}(\mathbb{T}^d)$
or in $KSp^{-n+d}(\mathbb{T}^d)$. Here, $p^*$ is the pull-back under the projection $p$, while $\widehat{p}_*$ is the push-forward in real $K$-theory \cite{Kar}. The latter can be viewed as a real index \cite{AS5} of elliptic operators along the fibres.

Physically, the groups on the right-hand-sides of \eqref{FM1}-\eqref{FM2} classify topological insulators in $d$-dimensions compatible with a subgroup of $\{1,T\}\times\{1,C\}$-symmetries which determines the value of $n$ as in Table \ref{table:0dgappedphasestable}. Similarly, the commutation relations amongst the set $\{i,\mathsf{P}\mathsf{T},\mathsf{P}\mathsf{C},\Gamma\}$ determine the Clifford algebra action that they generate, which we observe is now a \emph{fibrewise} action as explained in Subsection \ref{subsection:PCTinsulators}. Therefore, the groups on the left-hand-sides of \eqref{FM1}-\eqref{FM2} classify topological insulators in $d$-dimensions compatible with a subgroup of $\{1,TP\}\times\{1,CP\}$ symmetries.

In other words, there is a duality between topological insulators (superconductors) with $C$-$T$ symmetries and those with $CP$-$TP$ symmetries. For example, if we take $d=2$ and $n=4$, then $KR^{-4}(\widehat{\mathbb{T}^2})$ classifies fermionic $\mathsf{T}$-invariant insulators in two dimensions. The dual system has $\mathsf{C}\mathsf{P}$ symmetry squaring to $+1$, being described by the dual $K$-theory group $KO^{-2}(\mathbb{T}^2)$. In the special case $d=4$, the duality under the FM-transform involves the modification of symmetries $\mathsf{T}\leftrightarrow \mathsf{TP}$, $\mathsf{C}\leftrightarrow \mathsf{CP}$ along with an adjustment of their squares $\pm 1\mapsto \mp1$. For example, $KR^{-4}(\widehat{\mathbb{T}^4})$ describing $\mathsf{T}$-invariant insulators in 4D with $\mathsf{T}^2=-1$ is dual to $KO^0(\mathbb{T}^4)$ describing $\mathsf{T}\mathsf{P}$-invariant insulators with $(\mathsf{T}\mathsf{P})^2=+1$.

As an independent check, we have the following well-known isomorphisms of groups \cite{Kar,H},
\begin{subequations}
\begin{align}
KO^{-n}(\mathbb{T}^d)&\cong \bigoplus_{j=0}^d \left[KO^{-n- j}(\star)\right]^{d\choose j},\\
KR^{-n}(\mathbb{T}^d)&\cong \bigoplus_{j=0}^d \left[KO^{-n+ j}(\star)\right]^{d\choose j},
\end{align}
\end{subequations}
which is consistent with \eqref{FM1}-\eqref{FM2} above.

\section*{Acknowledgements}
This work was supported by the Australian Research Council  
via ARC Discovery Project grants DP110100072, DP150100008 and DP130103924. We also thank Peter Bouwknegt for feedback on an earlier version of the paper.


\end{document}